\documentclass{elsarticle}

\usepackage{lineno}
\modulolinenumbers[5]
\usepackage[left=2cm, right=3.2cm,left=3.2cm, top=2cm]{geometry}
\usepackage[hidelinks]{hyperref}

\usepackage{amsmath,graphicx}
\usepackage{epstopdf}
\usepackage{amsfonts}
\usepackage{algorithm}
\usepackage{algpseudocode}
\usepackage{indentfirst}
\usepackage{upgreek}

\usepackage{placeins}

\journal{XXXXXX}

\begin{document}

\begin{frontmatter}

\title{Arbitrarily large iterative tomographic reconstruction on multiple GPUs using the TIGRE toolbox}

\author[soton]{Ander Biguri\corref{mycorrespondingauthor}}
\cortext[mycorrespondingauthor]{Corresponding author}
\ead{ander.biguri@gmail.com}
\author[soton]{Reuben Lindroos}
\author[mircoen]{Robert Bryll}
\author[soton]{Hossein Towsyfyan}
\author[soton]{Hans Deyhle}
\author[muvis]{Richard Boardman}
\author[muvis]{Mark Mavrogordato}
\author[cern]{Manjit Dosanjh}
\author[cern]{Steven Hancock}
\author[soton,muvis]{Thomas Blumensath}

\address[soton]{Institute of Sound and Vibration Research (ISVR), University of Southampton, Southampton SO17 1BJ, UK}
\address[mircoen]{Micro Encoder Inc, USA}
\address[cern]{CERN, Meyrin 1211, Switzerland}
\address[muvis]{$\upmu$-VIS X-ray Imaging Centre, University of Southampton, Southampton SO17 1BJ, UK}

\begin{abstract}
Tomographic image sizes keep increasing over time and while the GPUs that compute the tomographic reconstruction are also increasing in memory size, they are not doing so fast enough to reconstruct the largest datasets. This problem is often solved by reconstructing data in large clusters of GPUs with enough devices to fit the measured X-ray projections and reconstructed volume. Often this requires tens of GPUs, which is a very economically expensive solution. Access to single-node machines designed to reconstruct using just one or a few GPUs is more common in the field, but current software does not allow iterative reconstruction of volumes that do not fit in those GPUs. In this work, we propose a strategy to execute efficiently the required operations for iterative reconstruction for arbitrarily large images with any number of GPUs with arbitrarily small memories in a single node. Strategies for both the forward and backprojection operators are presented, along with two regularization approaches that are easily generalized to other projection types or regularizers. The proposed improvement also accelerates reconstruction of smaller images on single or multiple GPUs, providing faster code for time-critical medical applications. The resulting algorithm has been added to the TIGRE toolbox, a repository for iterative reconstruction algorithms for general CT, but this memory-saving and problem-splitting strategy can be easily adapted for any other GPU-based CT code.
\end{abstract}


\end{frontmatter}


\section{Introduction}

In recent years image sizes in tomographic applications have increased significantly due to advancement of X-ray sources, detector technologies and computational resources, while at the same time the available reconstruction and processing time requirements are getting smaller. In computed tomography (CT) for industrial or scientific applications, scanners that can cover a detector area of 2 m $\times$ 3 m are already available commercially, with detector pixels sizes in the order of 0.2 mm, resulting in very high resolution projections and consequently very high resolution images. These images are a big challenge to reconstruct as they can be tens to hundreds of GiB in size, bigger than the internal memory of current Graphic Processing Units (GPUs), the devices that are most commonly used for reconstruction in CT. On the other hand, industrial and medical applications of CT are also pushing for faster reconstruction. CT for quality control and metrorology of manufactured components requires fast scan times as increased throughput of samples is desired\cite{Kruth2011821}\cite{warnett2016towards}. In medical applications, especially with adaptive and image guided radiation therapy, short scan and reconstruction times are fundamental for the correct targeting of tumours\cite{letourneau2005cone}\cite{hansen2018fast}. 

Particularly in the medical field, but also in other applications, iterative algorithms are becoming more common as they can provide higher quality images in low data quality and quantity scenarios, reducing scan times and radiation dose\cite{Desai2012}\cite{doi:10.1177/1533033818823054}\cite{kataria2018assessment}. However, iterative algorithms are significantly slower than standard filtered backprojection based reconstruction methods, working against the goal of fast computation times and larger image sizes. 

Reconstructing large images with parallel computation is hardly a new or unexplored problem, a large amount of literature is available on the topic, especially for the Filtered Backprojection (FBP) and its cone-beam corrected modification, the Feldkamp Davis Kress (FDK) algorithm. For the parallel beam case, the problem can be easily distributed in either GPUs or  High Performance Computers (HPCs) by splitting the problem into axial slices and solving these independently\cite{atwood2015high}\cite{melvin2008hpc}\cite{mirone2014pyhst2}\cite{pelt2016integration}. For cone beam tomography the image can not be trivially split and reconstructed in independent pieces, complicating the possible parallelization procedures. CPU-based methods are the most common approach taken for parallelization in this case, as they require no change in the code, they will work as long as they are executed with a sufficiently large CPU-based computer, such as a HPC cluster\cite{1466377}\cite{rosen2013iterative}\cite{kachelriess2006hyperfast}. However GPUs have significant advantages for CT reconstruction, and several single and multi-GPU approaches have been proposed\cite{palenstijn2015distributed}\cite{fehringer2014versatile}\cite{zhu2012multi}, often using a message passing protocol (MPI) \cite{palenstijn2017distributed} to allow execution in large clusters of nodes with a few GPU's each. This method is possibly the best for large image CT reconstruction as it can be highly efficient computationally if the partitions of the image are optimized to minimize memory transfer\cite{jimenez2012irregular}\cite{buurlage2019geometric}. 

The currently available literature and software however has a major drawback when reconstructing large images on multiple GPUs: the whole set of projections and images have to fit inside the GPUs internal RAM. To reconstruct an image volume with $2000^3$ voxels, this requires around 8 of the highest memory GPUs, whilst for $4000^3$ voxel volumes over 50 GPUs are required, depending on the projection size. Access to a computer with these resources is not common and would be extremelly expensive. Most industrial, scientific and medical centres do not have access to such multi-GPU HPC clusters and instead have dedicated single node machines for CT data processing, with a few GPUs each (with numbers ranging from 2 to 8 GPUs, at best). Some published GPU implementations of the FDK algorithm can overcome this issue by dividing the image into separate pieces and treating each one independently\cite{kaseberg2013opencl}\cite{7237019}, but a general method for arbitrarily large iterative reconstruction on arbitrarily small GPUs is not available. In this work we propose a strategy to partition the projection and backprojection operations for efficient single-CPU multi-GPU execution of large CT images.

Most multi-GPU tomographic reconstruction code splits the mathematical operators such that the standard reconstruction algorithms remain unmodified. However, an alternative approach would be to split the problem mathematically instead\cite{hamalainen2014total}\cite{sorensen2014multicore}\cite{gao2018joint}, proposing different, distributed optimization methods that use multiple locally converging independent computations that ensure global convergence of the reconstruction. Each of these threads can be executed in a separate machine with limited memory transfer between nodes. These methods can be executed in arbitrarily large computer networks, but use different algorithms to solve the reconstruction problem.

The rest of this paper is structured as follows. Firstly, in the methods section, an overview of TIGRE and iterative algorithms is given, together with a review of GPU terminology. The section continues to explain the proposed approach for the forward and backprojection operations and finishes with a description of how to accelerate regularization operations that are implemented on GPUs. We then present results on execution times for different systems with varying image and detector sizes, together with two examples reconstructed with measured data.

\section{Methods}

The strategies for efficient multi-GPU tomographic reconstruction are primarily based on rearranging computation and memory transfer operations in such a way that the amount of computation is minimized while the overlap between memory transfer and computation execution is maximized. 

The Tomographic Iterative GPU-based Reconstruction toolbox (TIGRE) is a MATLAB and Python based toolbox for GPU accelerated tomographic reconstruction\cite{biguri2016tigre}. It is especially focused on iterative algorithms and provides implementations of more than ten of them, together with the FBP and FDK algorithms.

Mathematically speaking, iterative algorithms solve the linearized model
\begin{equation}
Ax=b+\tilde{e} \label{eq:linearized model}
\end{equation}
 where $x\in \mathbb{R}^{N_{e}}$ is a vector of image elements, $b\in \mathbb{R}^{N_{p}} $ a vector of absorption measurements and $A$ a matrix that describes intersection lengths between rays and elements. There are multiple approaches in the literature to solve equation \ref{eq:linearized model} using iterative methods, however all of them have something in common: at least one of each $Ax$ (forward projection) and $A^Tb$ (backprojection) operations are required in each iteration.  These are the computationally heavy operations that take more than 99\% of the execution time. Some algorithms also contain other computationally heavy operations, such as regularization operations. These operations are implemented in TIGRE using GPU accelerated code for NVIDIA devices using the CUDA language. These core operations are the ones needing adaptation if arbitrarily large images are to be reconstructed. Inefficient strategies in splitting and gathering of the partial results can lead to unnecessary computations and memory transfers. Optimizing the splitting strategy for the general case is the goal of this article.

\begin{figure}

\begin{center}
\includegraphics[width=0.7\textwidth]{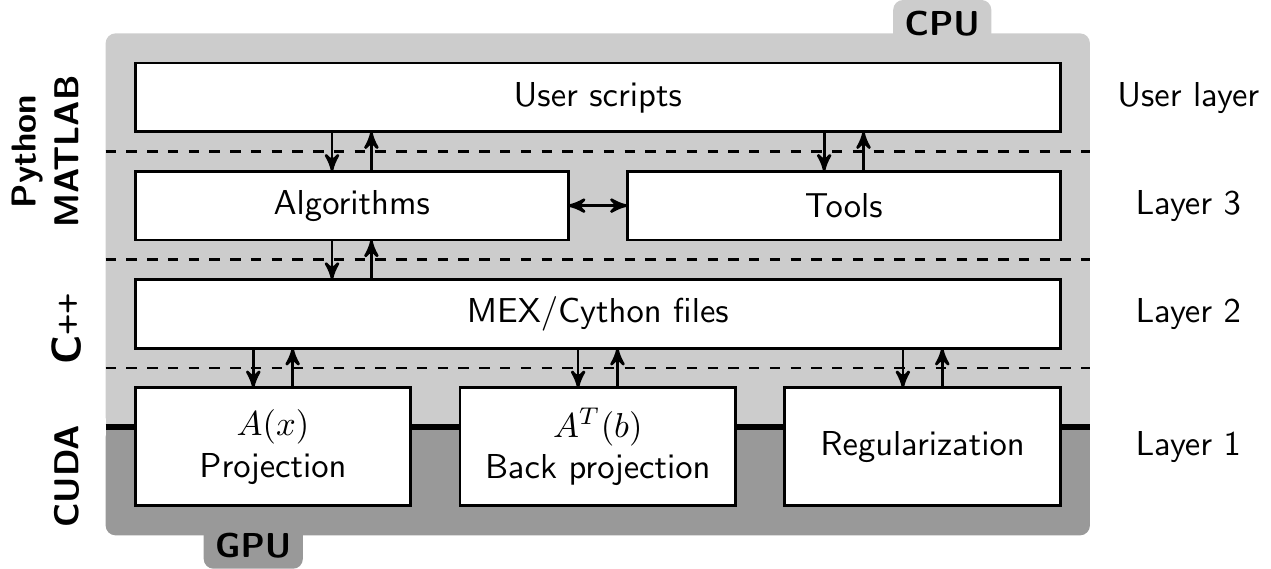}

\end{center}
\caption{Modular architecture of the TIGRE toolbox.}
\end{figure}

 TIGRE's architecture is modular, thus all of the GPU code is independent from the algorithm that uses it. Thus, by adapting the GPU code to be able to handle arbitrarily large reconstruction, TIGRE will also automatically handle such images. The modular design however has a drawback. To allow the higher level abstraction required for the design of new algorithms by turning the computationally heavy operations into ``black box'' functions, sacrifices are needed in memory management and algorithm specific optimizations.  Specifically, TIGRE will move memory from CPU to GPU and vice-versa in each GPU call. This allows for fast prototyping of new methods, but means algorithms may perform slightly slower than in fully optimized GPU code. The architecture also brings another limitation, especially for multi-GPU code: as memory is allocated and deallocated in the higher level languages (MATLAB and Python), it is stored in pageable memory. In brief, this is memory managed by the operating system (OS) that may be split, moved around, or not fully stored in RAM if the OS deems it necessary. An alternative would be page-locked or pinned memory, memory that as it's name suggest, is locked into a certain RAM location. This type of memory has not only faster transfer speeds (as it can be managed by the GPU without CPU intervention), but also allows for an important feature in speeding up code, asynchronous memory transfers.
 
CPU calls to GPU instructions can either be synchronous or asynchronous. If the call is synchronous, the CPU waits until the completion of the instruction before continuing, while if asynchronous, the CPU will keep executing further instructions, regardless if the GPU has terminated its task or not. Queueing operations but ensuring none of them are executed before its requirements are met (e.g. some computation finished, specific memory transfers ended) is the key feature for minimizing total computational time on multi-GPU systems, as it allows queuing simultaneous memory reads or writes and simultaneous memory management and computation execution. As TIGRE relies on Python and MATLAB to allocate this memory, memory will initially be paged thus only transferable synchronously, unless explicitly pinned beforehand, which is a costly operation. While the aforementioned limitations will cause the code presented here to be slightly less efficient, discussion on how to implement the multi-GPU strategy when the limitations imposed by TIGRE are not present is also given.

Before introducing the acceleration strategies used, it is worth mentioning that this work focuses on how to organize computation and memory transfer, but does not describe (nor constrain) the kernels (computational functions) themselves. There are several algorithms available in the literature\cite{chou2011fast}\cite{schlifske2016fast}\cite{okitsu2010high}\cite{long20103d} to implement forward and back-projectors efficiently on GPUs and fair comparison between codes is a challenging task, as code is often not available and execution times are hardware specific. We do not claim that the kernels used in this work are the fastest, but our multi-GPU strategy presented is applicable to most, if not all, the algorithms for forward and backprojection in the literature.

\subsection{Forward projection}

The forward projection operation, $Ax$, consists on simulating ideal X-ray attenuation with a given image estimate, $x$. There are several published algorithms to approximate the operation, two of them which are implemented in TIGRE: interpolated projection\cite{chou2011fast} and ray-driven intersection projection with Siddon's method\cite{siddon1985fast}. While the integral over the X-ray paths is computed differently for each methods, the kernels are organized and executed using the same structure.

The kernels are queued and executed on GPUs using 3D blocks of threads of size $N_u\times N_v\times N_{angles}$\footnote{For NVIDIA GTX 10XX GPUs, values of $N_u=9$, $N_v=9$, $N_{angles}=9$ have been found to be the fastest empirically. Different values do not alter the algorithms proposed in this work.}, where each thread processes the integral of an individual detector pixel, as seen in Figure 2. By using the neighbouring memory caching capabilities of the texture in the GPU, this kernel structure allows for faster executions due to higher memory read latency. This is because memory reads needs hundreds of cycles to complete, but storing the image in texture memory allows caching in 3D. Therefore concurrent threads reading in the same or neighbouring elements cause an increased cache hit rate, directly translated into lower memory read time and faster execution. On top of the faster memory reads, texture memory also allows for hardware implemented trilinear interpolation. For more information on the projection approach, the reader is reffered to the technical note by Palenstijn \textit{et al}\cite{palenstijn2011performance}. Each time the forward projection kernel is launched enough thread-blocks to cover the entire height and width of the projection are queued, but only one in depth. Thus, each kernel launch will compute $N_{angles}$ entire projections.

\begin{figure}[ht]
\begin{center}
\includegraphics[width=0.6\textwidth]{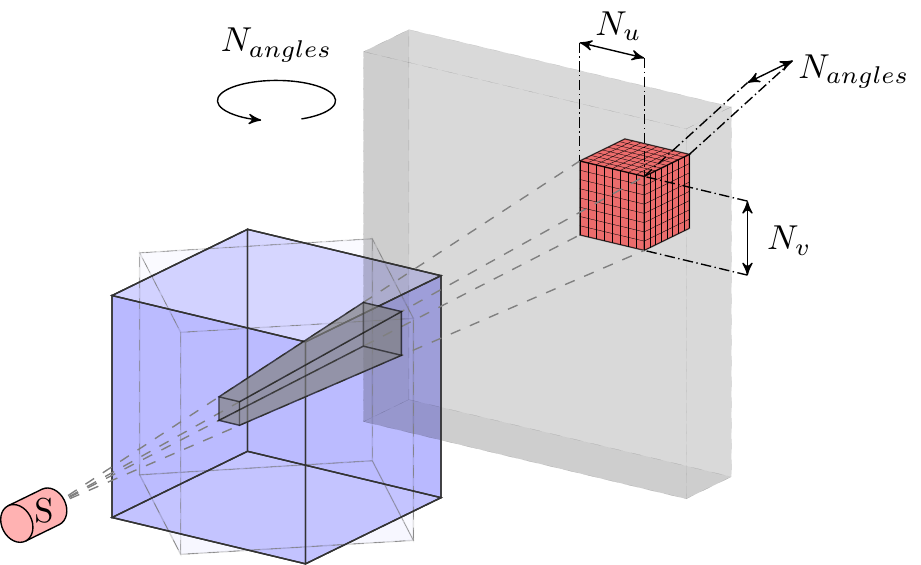}
\end{center}
\label{fig:fP_schematics}
\caption{The forward projection block execution kernel structure. The detector is divided in $N_u\times N_v\times N_{angles}$ sized blocks to maximize cache hit rates. {Each kernel launch executes  enough blocks to fully compute $N_{angles}$ entire projections.}}

\end{figure}

To execute the projection operation in cone beam geometries (such as circular, C-arm or helical scans, among others) however, the entire image is required, as due to the shape of the X-rays different image partitions will have influence in overlapping areas in the detector. Thus if the image does not fit in the GPU, projections will need to be separately computed for each image partition and later accumulated. The accumulation procedure is ultra-fast, executing in approximately 0.01\% of the time that a projection kernel launch needs to finish.

A common approach for a projection operation on multiple GPUs is to allocate the memory of the projections along with the image size, then execute the kernels. Then a gathering step is performed to accumulate the overlapping detector pieces. This is a good approach if the image and projections fit in the available GPU memory, but for large images this is not the case thus the variables needs to be partially stored in the host RAM. Once this becomes necessary, then it is important to minimize the amount of unnecessary memory on the GPUs, in order to maximize the size of each individual image partition (i.e. minimize the amount of splits to the image).

The approach we propose can be summarised as follows: Only the memory for two projection kernel launches will be allocated in each device ($2\times N_{angles}$ individual projections) and the rest of the GPU memory will be free for the image. These two projection pieces will act as a buffer to store the computed projections while they are being copied out to the CPU during the kernel execution. As one piece is being used to store the projections that are being computed at that moment, the other will be copying memory to the CPU RAM. This not only allows us to minimize the number of splits needed, it also reduces execution times, as no extra time is needed to copy the results out of the GPU. On multi-GPU systems, each GPU will compute a set of independent projections, using the entire image. If the image does not fit entirely in each of the GPUs, the image is partitioned into same size volumetric axial slices stacks, as big as possible. Additionally, an extra projection buffer will be allocated with the same size. These will be used after the first image slice has been forward projected, as they will load previously computed partial set of projection data into GPU RAM, to be accumulated on the GPU when each of the projection kernels are finished.

Queueing these operations in the correct order is key, as memory copies will halt the CPU code until completion, impeding execution of concurrent operations. Once the image slice is already in the GPU, first the kernel launch for the projection is executed asynchronously. Then, if needed, the partial projections from previously computed image slices are copied from CPU to each of the GPUs. After the completion of the copy, the CPU queues asynchronously the accumulation kernel and finally copies the result from previous computation to CPU. Each of these instructions is executed for all available GPUs simultaneously, as opposed to executing all the instructions on one GPU at a time. A timeline of the process can be seen in figure 3 and Algorithm \ref{alg:proj} shows the pseudocode.   

In the one and two GPU case, empirical tests show that if the image needs to be partitioned due to lack of memory, page-locking the image memory at the beginning of the process results in a faster overall execution. This is true even with the significant time it requires to page-lock, thanks to faster CPU-GPU transfer speeds (from approximately 4GB/s to 12GB/s on a PCI-e Gen3). On more than two GPUs, page locking always brings an improvement, due to the simultaneous copy that it enables.

\begin{algorithm}
    \caption{Projection operation kernel launch procedure}
    \label{alg:proj}
    \begin{algorithmic}[1] 
    \State Check GPU memory and properties
    \State Split projections among GPUs
    \If {Image does not fit on GPU}
    \State Page-lock image memory  \Comment{If the device allows it}
    \EndIf
    \State Initialize required synchronous operations (texture, buffers, auxiliary variables)
    \For {$N_{sp}$ image splits}
        \State Copy current image split CPU$\rightarrow$GPUs \Comment{Asynchronous}
        \State \textbf{Synchronize}()
        \For{(Total angles)$/N_{GPU}/N_{angles}$ amount of kernel calls}
        	\State ForwardProject$<<<$\textbf{Launch}$>>>$ \Comment{Asynchronous}
        	\If{Not first image split}
        	    \State Copy already computed partial projections CPU$\rightarrow$GPUs \Comment{Synchronous}
        	    \State \textbf{Synchronize}(Memory) \Comment{Wait until memory copy completion}
        	    \State AccumulateProjections$<<<$\textbf{Launch}$>>>$\Comment{Asynchronous}
            \EndIf
        	\If{Not first projection}
        	    \State Copy previous kernel projections GPUs$\rightarrow$CPU \Comment{Synchronous}
        	\EndIf
        	\State \textbf{Synchronize}(Compute) 
        \EndFor 
         \State Copy last kernel projections GPUs$\rightarrow$CPU\Comment{Synchronous}
         \State \textbf{Synchronize}()
	\EndFor
	\State Free GPU resources  
    \end{algorithmic}
\end{algorithm}

\begin{figure}

\begin{center}
\includegraphics[width=1\textwidth]{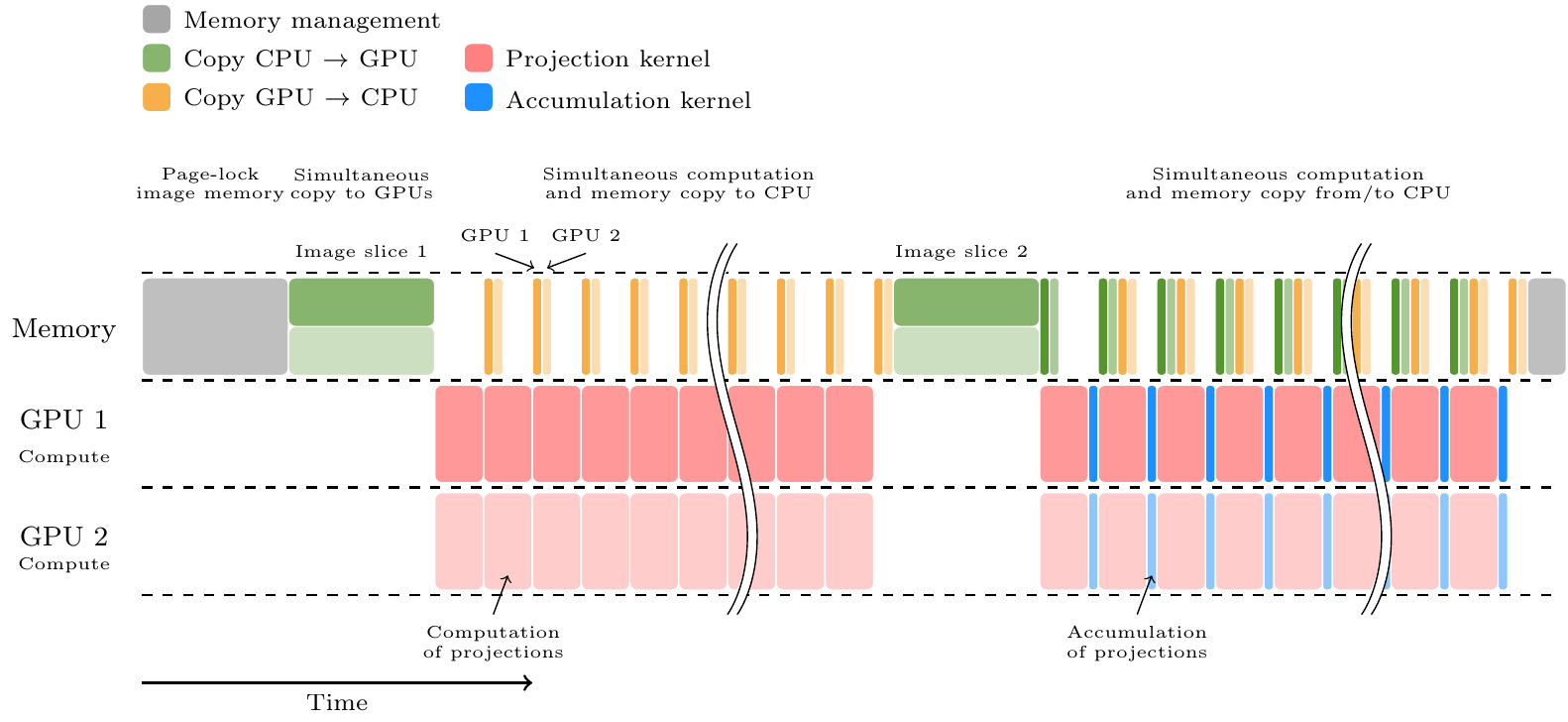}

\end{center}
\label{fig:FPtimeline}

\caption{Forward projection timeline for 2 GPUs and an image that requires 2 splits per GPU (not to scale). Each GPU handles half of the projections required. Using buffers, projection results are asynchronously copied to CPU while the next kernel launch is being computed. On image partitions other than the first, copies of the partial projections are copied while the current set is being computed for accumulation. }

\end{figure}

\subsection{Backprojection}
\begin{figure}[ht]

\begin{center}
\includegraphics[width=0.6\textwidth]{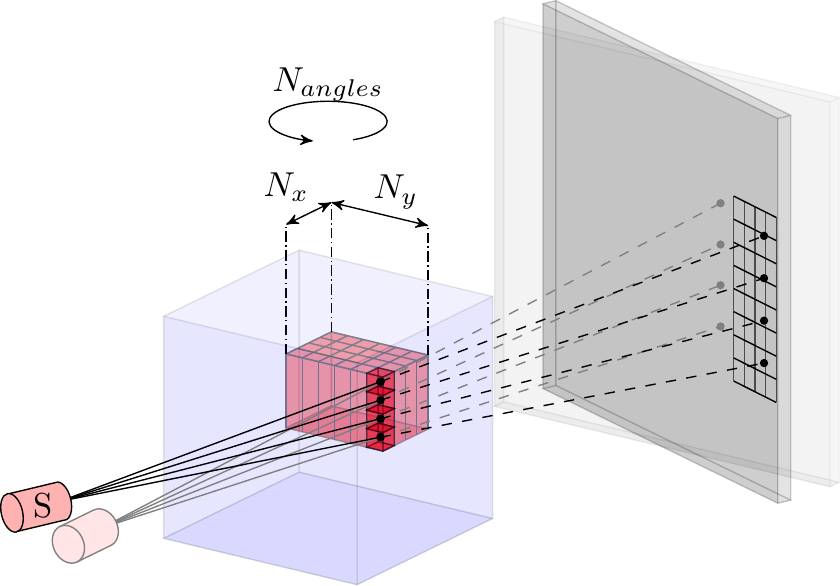}

\end{center}
\caption{Back-pojection kernel structure for a single thread-block. Each kernel launch executes a set of blocks covering $N_{angles}$ projections using the entire image.}
\label{fig:BP schematics}
\end{figure}

The backprojection operation, $A^Tb$, smears detector values, $b$ along the X-ray paths towards the source. TIGRE uses a voxel-based backprojection, with two optional weights, FDK weights and ``pseudo-matched'' weights that approximate the adjoint of the ray-driven intersection projector\cite{jia2011gpu}. Both backprojectors execute $N_x \times N_y \times N_{angles}$ block of threads, where each thread computes $N_z$ voxel updates\footnote{In TIGRE $N_x=16$, $N_y=32$, $N_{angles}=32$ and $N_z=8$. Different values do not alter the algorithms proposed in this work.}, as described in detail in articles\cite{papenhausen2011gpu} and \cite{zinsser2013systematic} and shown in figure \ref{fig:BP schematics}. Similar to the projection operators, this execution order increases occupancy of the processors and optimizes cache hit rates, reducing the overall computational times. Therefore the projections need to be stored in texture memory to enable 3D caching.

Similarly to the projection, the common approach for back-projecting in the literature requires the allocation of the entire set of projections on GPU RAM before or during the computation of the backprojection voxel updates. This approach does not minimize memory usage. 

\begin{figure}

\begin{center}
\includegraphics[width=\textwidth]{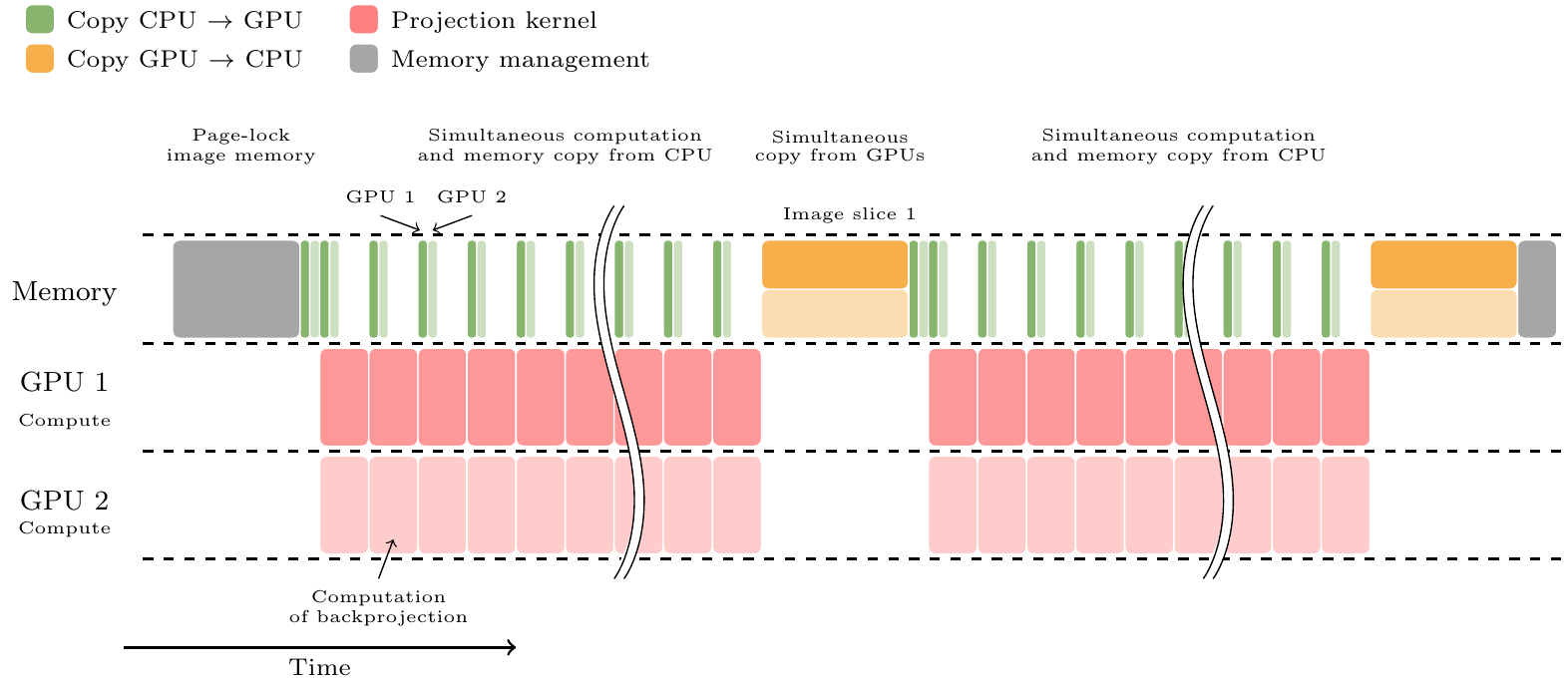}

\end{center}
\caption{Backprojection timeline for 2 GPUs and an image size that requires 2 splits per GPU (not to scale). Each GPU will update each corresponding image slice using the entire set of projections. These are copied to GPU while the backprojection is being computed to minimize transfer times.}
\label{fig:BPtimeline}
\end{figure}

The proposed algorithm for multi-GPU splitting is presented in algorithm \ref{alg:back}. Not unlike in the projection operation, two buffers of size $N_{angles}$ are allocated in each GPU for the projections. Before computation starts, the image is split into equal sized volumetric axial slices stacks and allocated among GPUs. If the total image (plus buffers) does not fit in the total amount of GPU RAM among devices a queue of image pieces is added and each GPU is thus responsible of back-projecting more than one image slice. Each GPU uses all the projections, but performs the CPU-GPU memory transfer simultaneously with the voxel update computation. Unless the image size is considerably smaller than the detector size (which does not happen in real datasets) the memory transfer should complete sufficiently fast. The execution timeline can be seen in figure \ref{fig:BPtimeline}. Empirical tests show that page-locking the image memory is faster than using pageable memory when a single GPU needs to compute multiple image pieces.
 
\begin{algorithm}
    \caption{Backprojection operation kernel launch procedure}
    \label{alg:back}
    \begin{algorithmic}[1] 
    \State Check GPU memory and properties
    \State Split projections among GPUs
    \If {Image does not fit on GPU}
    \State Page-lock image memory  \Comment{If the device allows it}
    \EndIf
    \State Initialize required synchronous operations (texture, buffers, auxiliary variables)
    \For {$N_{sp}$ image splits}
        \For{(Total angles)$/N_{angles}$ amount of kernel calls}
            \State Copy projection set CPU$\rightarrow$GPUs
            \State \textbf{Synchronize}()
        	\State Backproject$<<<$\textbf{Launch}$>>>$ \Comment{Asynchronous, it gets queued} 
        \EndFor 
        \State Copy image piece GPUs$\rightarrow$CPU
	\EndFor
	\State Free GPU resources  
    \end{algorithmic}
\end{algorithm}

\subsection{Regularization}
Tomographic reconstruction algorithms may include prior information to enforce a user defined constraint. While a variety of regularizers are available in the literature\cite{chen2008prior}\cite{semerci2014tensor}\cite{piccolomini1999conjugate}, the total variation (TV) regularization is arguably the most common in CT algorithms\cite{sidky2008image}\cite{lohvithee2017parameter}. Therefore we focus here on TV type constraints. The logic used to speed up and split the problems however can be applied to other forms of neighbourhood based regularizers.

TV regularization adds an additional constraint to the image reconstruction problem: sparseness of the gradient, enforcing piecewise flat images. 
Two distinct formats of minimizing the TV constraint are used in TIGRE that appear in different algorithms in the literature: The gradient descend minimization approach and the Rudin-Osher-Fatemi (ROF) model minimization approach\cite{rudin1992nonlinear}\cite{knoll2010fast}. Both of the codes use different minimization algorithms to reduce the total variation, but have two things in common. First, both methods take an image volume as input and return an image volume of the same size, i.e. the algorithms are not specific to CT. Secondly, similar to some other regularizers, they require knowledge of adjacent voxels to update each voxel value, thus are coupled operations. This makes them considerably harder to parallelise on multiple devices, as synchronization of the voxel values is required in each iteration, requiring communication between GPUs. Both methods often need tens or hundreds of iterations to converge.

\begin{figure}[ht]

\begin{center}
\includegraphics[width=0.6\textwidth]{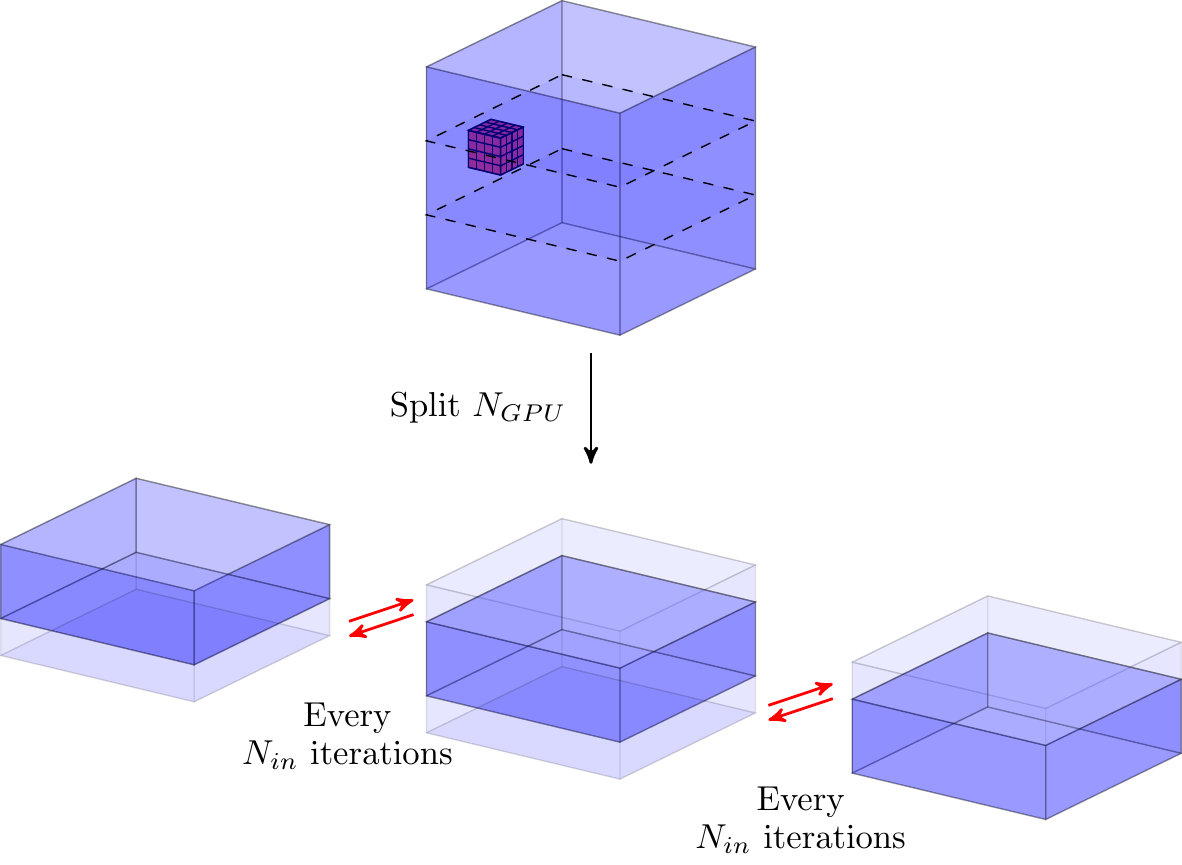}

\end{center}
\caption{Regularization kernel diagram and splitting procedure for $N_{GPU}=3$. Each kernel launch updates the entire image with multiple blocks. Each split contains $N_{in}$ deep buffers in its boundaries, allowing for $N_{in}$ independent inner iterations, before synchronizing the buffers.}
\label{fig:TVdiagram}
\end{figure}

While optimization of each of these regularizers is algorithm specific, the splitting method is the same as described in figure \ref{fig:TVdiagram}. An interesting property of this type of single voxel neighbouring is that if an overlapping buffer is created on the boundaries of the image containing voxel values of adjacent image splits, the depth of the buffer is equal to the amount of independent iterations that image piece can perform. This means that one can run stacks of $N_{in}$ amount of independent iterations of the minimization of the regularization function to then update the buffer values with the newly computed voxel values on other GPUs. However, the depth on the buffer directly translates to higher computational demand, as each GPU will need to repeat operations in the same slices. In our work, a depth value of $N_{in}=60$ in each split boundary has been found to have the best balance between minimizing memory transfer steps and computational time. Additionally, the algorithms may need a norm in each iteration. If an operation where the entire image is required is present, algorithm specific decisions need to be made. In the TV case, the norm of the image update is required in each iteration. Empirical measurements suggest that not synchronizing in each iteration and approximating the norm assuming uniform distribution along the image samples has negligible effect in the convergence and result of the algorithm.   

This approach however suffers heavily whenever the image and auxiliary variables do not fit entirely on the total amount of GPU RAM across devices. As quite often the regularization optimizers need several copies of the input image (e.g. the ROF minimizer in TIGRE requires 5 copies) the upper bound of the GPU RAM is easily reached. If this is the case, image splits need to be continuously read and written into CPU RAM, which heavily hinders overall performance. To maximize the data transfer speeds in these cases the memory is allocated and pinned in the CPU RAM. In the case where the copies fit in GPU memory, the buffers are also allocated and pinned in the host, but are of much smaller size.

\section{Results}

Two experiments are conducted to evaluate the proposed multi-GPU code. First, a detailed experiment is executed on the projection and backprojection operations themselves, using varying problem sizes and number of GPUs. Analysis of the speed-up and performance achieved in the different situations is given. Second, two large scans at different scales and from different X-ray machines are reconstructed using the multi-GPU code and the TIGRE toolbox. 

\subsection{GPU call performance analysis}
To test the performance of the multi-GPU code, the CUDA code is executed with varying image sizes on up to 4 GPUs. Two machines are used for the experiments with very similar specifications. For the 1 and 2 GPU nodes, a workstation with a AMD EPYC 7551P processor with 256 GB of RAM and 2 NVIDIA GTX 1080Ti with 11 GiB of RAM each is used, connected with independent PCI-e Gen3 x16 ports. For the 3 and 4 GPU tests a GPU node on the Iridis 5 HPC cluster has been used, with 2 Intel Xeon E5 2680 processors, 128 GB of RAM and 4 NVIDIA GTX 1080 Ti connected with independent PCE-e Gen3 x16 ports. Different machines were used simply due to the larger RAM on the 2 GPU workstation. While the processors are different, empirical measurements showed no significant time difference when executing the 1 and 2 GPU code in the HPC GPU node, thus results are comparable.

\begin{figure}[ht]

\begin{center}
\includegraphics[width=0.9\textwidth]{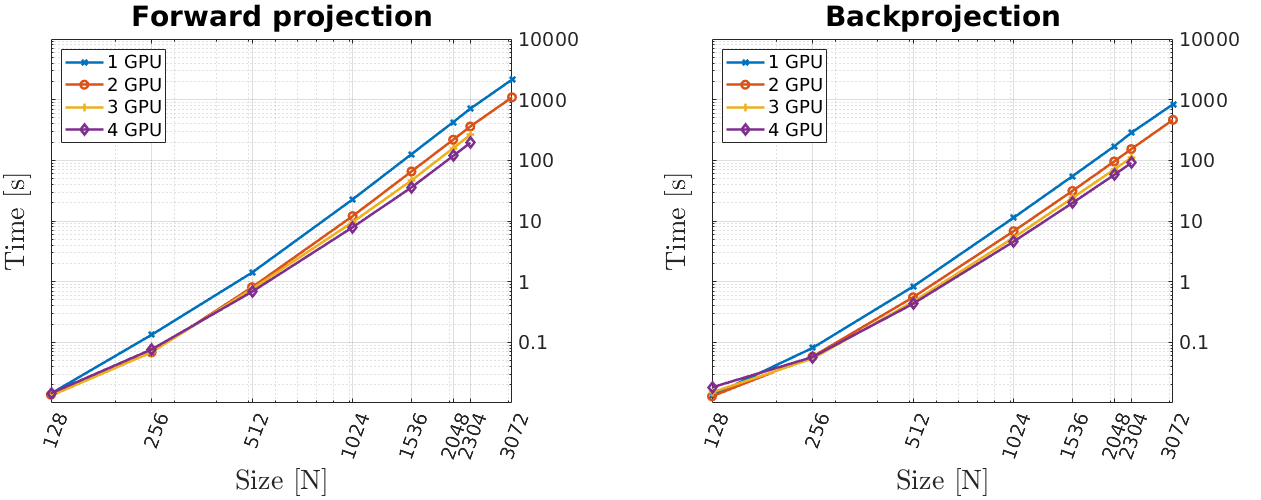}

\end{center}
\caption{Projection and backprojection speeds at different sizes ($N^3$ voxel volumes with $N^2$ projections at $N$ angles) for different number of GPUs. The time includes memory transfer speeds from and to the GPU. The missing points are due to lack of CPU RAM in the 4 GPU machine.}
\label{fig:speed}
\end{figure}

\begin{figure}[ht]

\begin{center}
\includegraphics[width=0.9\textwidth]{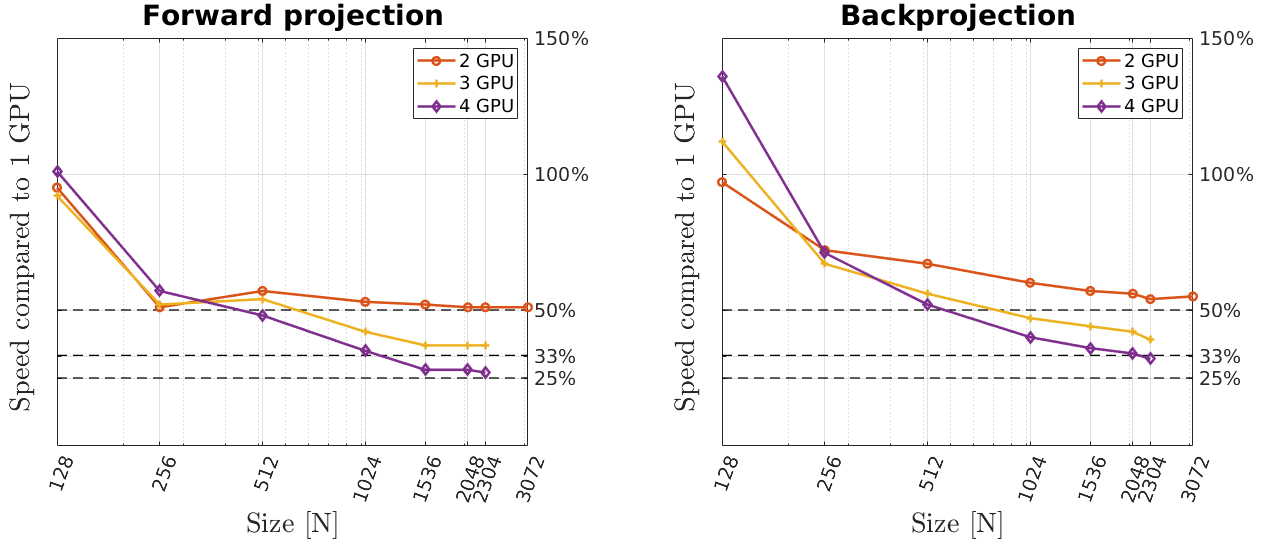}

\end{center}
\caption{Projection and backprojection computational time percentages compared to the 1 GPU execution time at different sizes ($N^3$ voxel volumes with $N^2$ projections at $N$ angles) for different number of GPUs. The time includes memory transfer speeds from and to the GPU. The missing points are due to lack of CPU RAM in the 4 GPU machine.}
\label{fig:raio}
\end{figure}

Figure \ref{fig:speed} presents the total time (computational plus memory transfer) for different sizes ($N$) and number of GPUs. The experiments used $N^3$ image volumes with $N^2$ detector pixels and $N$ projection angles each. Figure \ref{fig:raio} shows the percentage of time required compared to the single GPU code. TIGRE has two projection and backprojection types, but here we only time the default ray-voxel intersection algorithm for projection and FDK weighted backprojection. The alternative interpolated projector is not invoked by any of TIGRE’s algorithms because it is slower than the ray-voxel version. It was included in the toolbox for completeness and, in tests, gave virtually the same results. The backprojection with matched weights is 10\%-20\% slower than the FDK weighted ones and its only used when a matched backprojection is fundamentally required, for example when using the CGLS or FISTA algorithms. Apart from the slower kernels, the multi-GPU implementation is the same and thus their performance results are not shown.

The results show the expected performance gains. Computationally, duplicating the amount of GPUs should halve the kernel time. This is mostly true, the only difference is due to the last computational block in each kernel having less than their corresponding value of $N_{angles}$ and the extra computational time required for projection accumulation. The other reason for not reaching the theoretical speed increase is due to memory management and transfer times. At small problem sizes, the kernels are fast enough so that the time is dominated by GPU property checking and memory transfer. The computation kernels are small enough as to hide any improvement that multiple GPUs could provide. The special case of the backprojection parallelizing a size of $N=128$ results in a slower execution, but the overall time is smaller than 20 milliseconds. As $N$ gets larger, the computational cost of both operations increases, therefore the measured speedup ratios approach the theoretical ones (50\%, 33\% and 25\% for 2,3 and 4 GPUs respectively). This can be seen more clearly for the forward projection.  Adding more GPUs is generally expected to keep the ratio closer to the theoretical limit until there are enough GPUs with each having too little computational load compared to the memory transfer required, as happens with the smallest sizes on this experiment\footnote{Assuming a unique PCI-e Gen3 for each, otherwise the slower memory transfer time would impact the speed-up.}.

The backprojection doesn't scale as well as the projection for two main reasons. Firstly, it is faster. This means that there is a significantly larger influence of memory management on the overall execution time. Additionally, the backprojection requires pinning image memory which adds significant overhead. In the projection operation, the memory is reserved before execution, but the OS allocates it as it gets written.

 As an example of the splitting procedure, for the size $N=3072$, the single GPU node required 11 image partitions while the 2 GPU version required 6 partitions for the backprojection. The projection just needed 10 and 5 partitions each. The difference is due to the smaller value of $N_{angles}$ that the projection uses.

\begin{figure}[ht]

\begin{center}
\includegraphics[width=0.9\textwidth]{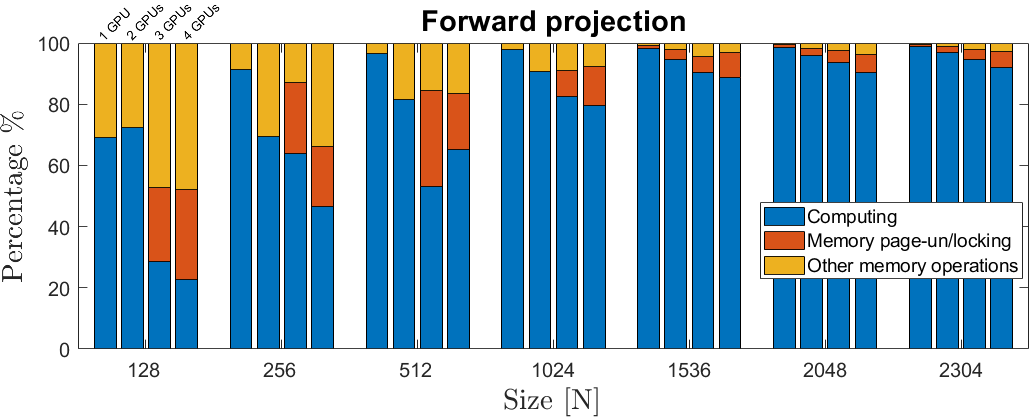}
\includegraphics[width=0.9\textwidth]{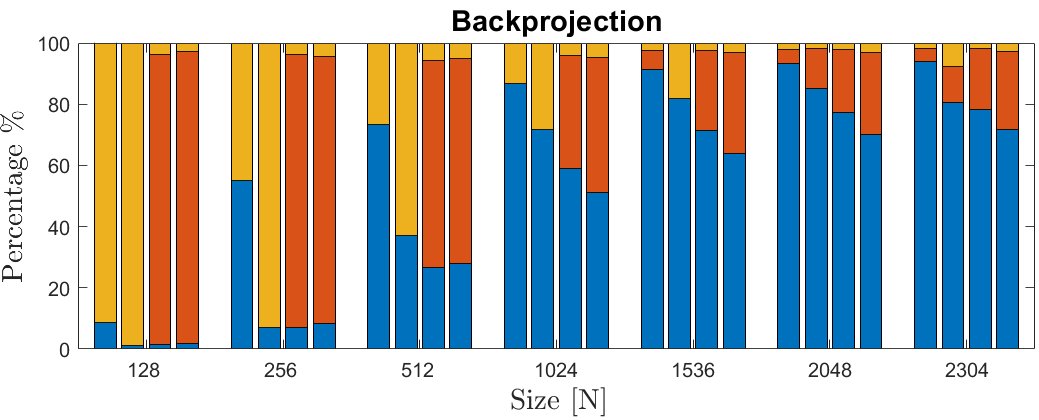}

\end{center}
\caption{Percentage of the total execution times by different operations compared to the image size on different amount of GPUs. Operations are grouped into three labels: Computing, memory page-locking and unlocking and other memory operations, such as allocating, freeing and memory transfer operations that are not concurrent with execution.}
\label{fig:pct}
\end{figure}

As this work proposes memory handling optimizations for GPU acceleration, measuring the impact that the memory has on the total execution of the code is of particular interest. Figure \ref{fig:pct} shows the result of different sizes and number of GPUs, for both the projection and backprojection. The operations are binned into three categories. Computing contains the time for kernel launches, which includes simultaneous memory copies as they happen concurrently. Then the time to pin and unpin the CPU memory is added. In the backprojection operation, this time is a bigger part of the total because it forces the CPU to allocate the memory, while in the forward projection, the memory is already allocated as the image volume already exists. Some problem sizes do not have this section as, due to page-locking being a slow operation, it is not performed because it brings less improvement than its cost. Finally, there are other memory operations. These include the time when the image is being copied in or out of the GPU and no computations are happening. It also includes memory freeing and other minor memory operations. 

This analysis also helps explain better Figure \ref{fig:pct}. One can see that in the forward projection the computing time dominates most of the execution times even for very small images, while in the backprojection, even at $512^3$ voxel volumes, the computation takes less than half of the time with more than 1 GPU.

\subsection{CT Image reconstruction}
In order to showcase the capabilities of the toolbox with the multi-GPU acceleration two images from two different scanners and at different resolutions are reconstructed. The computer used is the previously described dual-GPU machine with 256 GB of RAM. The first scan is of a roasted coffee bean and the second of a piece of a fossilized species of Ichthyosaur.

\subsubsection*{Coffee Bean on Zeiss Xradia Versa 510}
The first sample is of a roasted coffee bean. It was scanned on a Zeiss Xradia Versa 510, at \mbox{80 kVp}, 87.5 $\upmu$A (7 W). No filtration of the X-ray source was used. The source to object distance was 16mm with a geometric magnification of 9.48 and a macro objective of 0.4 magnification. Using a detector of 2000 x 2000 pixels a panel shifted scan was measured, offsetting the detector to the left and right to later stitch together into a 4000 x 2000 projection. Each shift was acquired at 6401 equally spaced rotation angles over 360 degrees, with 14 s of exposure, resulting in a 45 h scan. With these scan parameters, a voxel size of 3.653 $\upmu$m was achieved.

This scan was chosen to be reconstructed using the CGLS algorithm. As the algorithm needs several copies of the size of the reconstructed image and projections the entire dataset of 4000 $\times$ 4000 $\times$ 2000 voxels could not be reconstructed whilst keeping all data in system RAM. Therefore the projections were cropped into 900 $\times$ 3780 sized projections and only 2134 angles (a third) were used for reconstruction. This leads to a projection set that requires 29 GB of memory. An image of 3340 $\times$ 3340 $\times$ 900 voxels was reconstructed of size 12.2 $\times$ 12.2 $\times$ 3.28 $mm^3$, requiring 40 GB of memory. As previously mentioned the 2 GPUs on the reconstructed machine have 22 Gib or RAM between the two of them. 

\begin{figure}[ht]

\begin{center}
\includegraphics[width=0.95\textwidth]{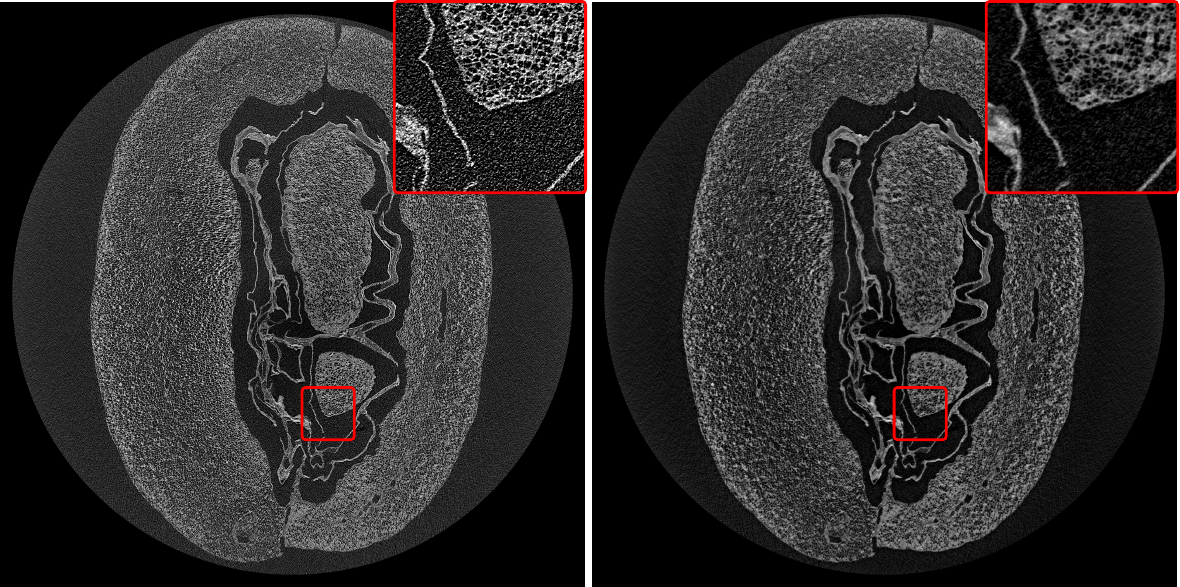}

\end{center}
\caption{FDK (left) and CGLS with 30 iterations (right) reconstructed slices of image sized 3340$\times$3340$\times$900 voxels of a roasted coffee bean. Reconstruction was performed on a 2 GPU machine.}
\label{fig:coffee}
\end{figure}

Figure \ref{fig:coffee} shows a slice of the reconstruction using the FDK algorithm (left) and the CGLS algorithm after 30 iterations. The CGLS reconstruction required 4 hours and 21 minutes. While at full angular sampling the FDK algorithm provides a high quality image, it starts showing noise artefacts over the reconstructed images when using a third of angles. The CGLS reconstruction is more robust against this noise. 

\subsubsection*{Ichthyosaur fossil on a Nikon/Metris 225kVp/450kVp custom bay}

The dataset used for this large image reconstruction is a small part of a fossilized Ichthyosaur species\cite{BBC}, the tip of the fins specifically. It was scanned on a Nikon/Metris 225 kVp/450 kVp custom bay, at 360 kVp, 3.37 $\upmu$A. Filtering of the source was performed using a 4 mm copper filter. The distance between the source and the object was 1.564 m, and the distance between the source and detector 2 m. The detector used had 2000 x 2000 pixels and was offset to the left and right of the sample during scan, giving a 0.8$\times$0.4 $m^2$ effective detector after image fusion. Sampling was performed uniformly at 6400 equaly spaced angular intervals. The image was reconstructed with a size of 3360 $\times$ 900 $\times$ 2000 with an isotropic voxel size of 0.156 mm.

\begin{figure}[ht]

\begin{center}
\includegraphics[width=0.31\textwidth]{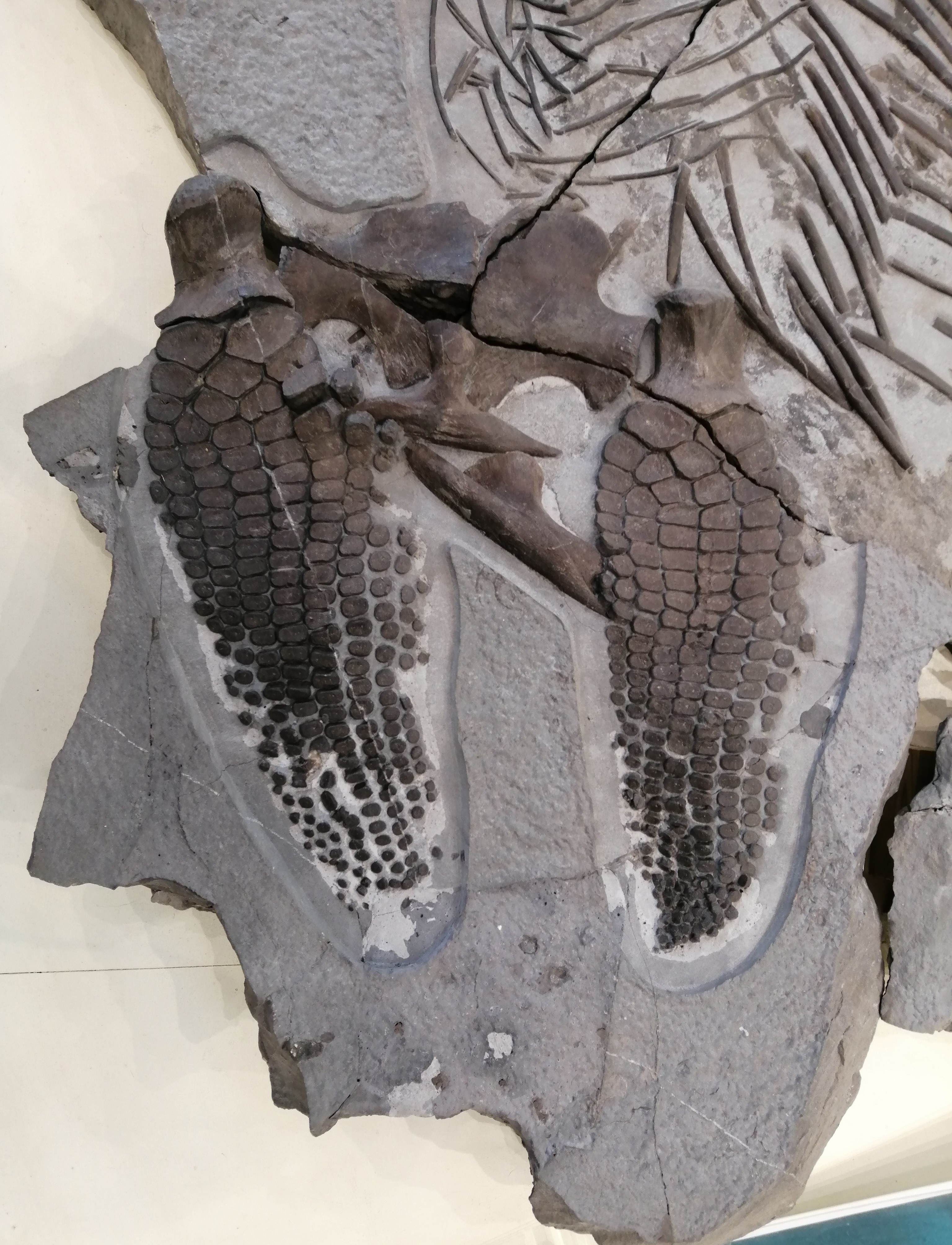} 
\includegraphics[width=0.64\textwidth]{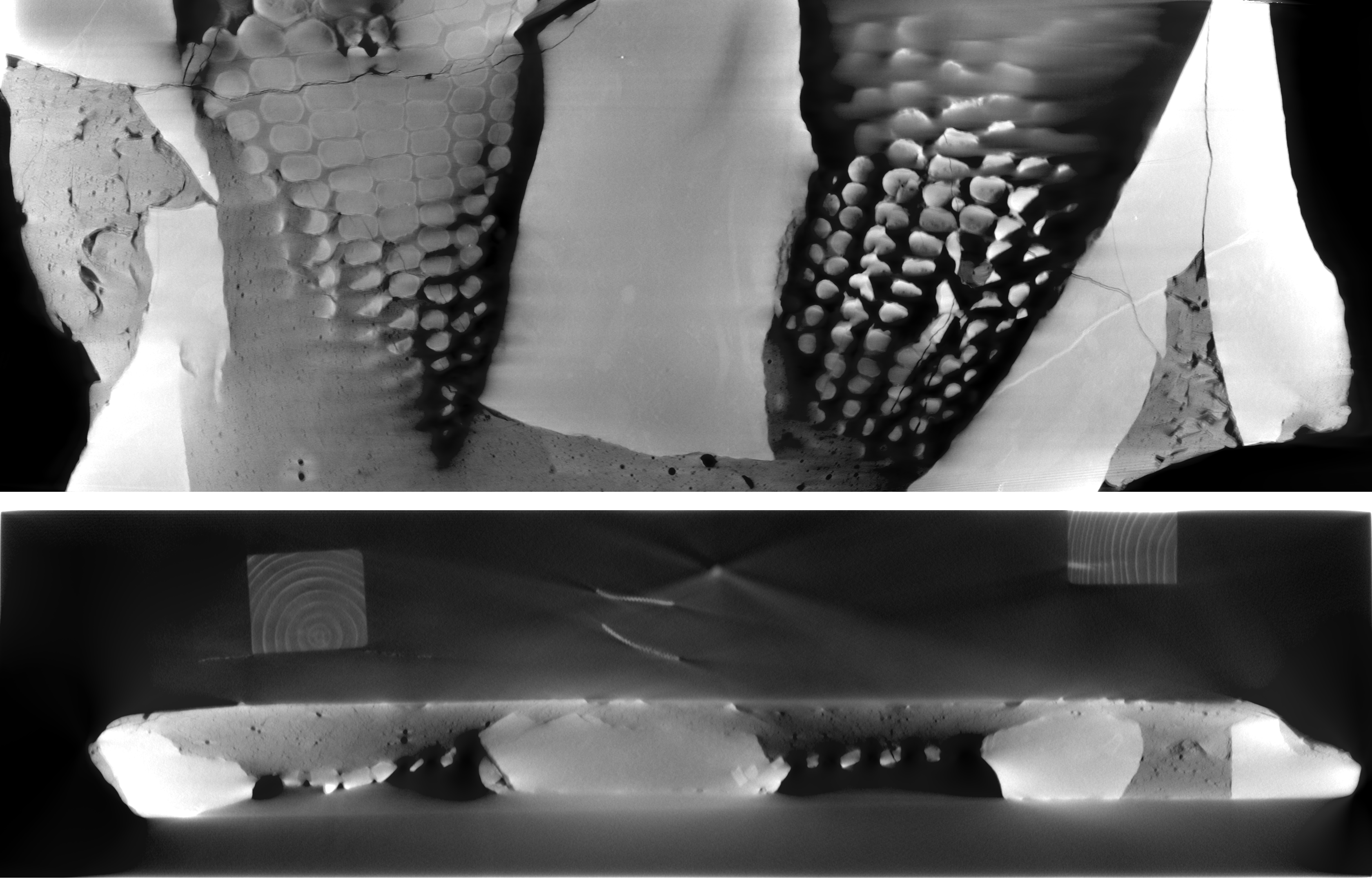}

\end{center}
\caption{Photography of the Ichthyosaur fossil (left) and OS-SART reconstruction with 50 iterations (right) reconstructed slices of image sized 3360$\times$900$\times$2000 voxels. Reconstruction was performed on a 2 GPU machine.}
\label{fig:dino}
\end{figure}

The reconstruction was performed using the OS-SART algorithm, a variation of the classic SART algorithm that updates the volume using partial projection subsets instead of updating it for each projection. The subset size was 200 projections and it was executed using 50 iterations, taking 6 hours and 40 minutes of execution time. Only 2000 uniformly sampled projections were used, in order to keep all data in system RAM. The image is 14.5 GB and the projections subset used 62 GB. Figure \ref{fig:dino} shows the specimen and slices from the reconstructed image.

\section{Discussion}

The results present a clear picture: with the proposed kernel launch structure very large images can be easily  reconstructed in parallel using multiple GPUs, minimizing the amount of image partitions needed. They also show how using multiple GPUs accelerates the reconstruction almost linearly, especially for bigger images. There are, however, some possible improvements that can be added into the presented structure and code. 

The simultaneous memory transfer and computation queueing strategy is the improvement that allows the multi-GPU code in this work to achieve high parallelism with little overhead, as there is no need to wait for projections to be copied in or out of the GPUs before or after the computation. However, in the work by Zinsser \textit{et al}\cite{zinsser2013systematic} a strategy that can be applied to overlap the image memory transfer is presented. In their work, the projections are entirely kept in the GPU memory but are partially copied in pieces of incremental size while their computation is being performed. Similarly, they divide the image into two slices in a single GPU, and start copying one of the pieces out to the CPU while the other one finishes computing. The splitting of the image into various slices per GPU can also be applied in the projection operation. This would allow in both operators to slightly decrease the memory transfer times. However, to ensure simultaneous compute and memory transfer the image needs to be split in very computer- and problem-specific sizes. Memory transfer speeds and different projection/image sizes are the main variables that influence how many splits the image would require. As TIGRE attempts to be as flexible as possible, this improvement has not been implemented, but it is noted here, as application-specific implementation of CT algorithms may want to consider it. The problem of memory operations possibly taking longer than kernels does not exist when handling projections, as the memory transfer times are always significantly faster than the computation of the kernels, unless an unrealistically big difference exists between the projection and image sizes. 

Some improvement can be achieved in the GPU code, especially in the backprojection, if an already preallocated memory piece can be given to the code. Due to the modular design of the toolbox, each kernel call allocates memory for its output, but due to the nature of iterative algorithms, it is likely that a variable with memory to overwrite already exists. We decided not to include this improvement and keep the modular design of TIGRE as it was.

The upper bound for memory is set by the CPU RAM available, however the GPU codes have their limitation. If the values for the fastest execution of kernels are respected ($N_{angles}$ in the projection operation, $N_{angles}$ and $N_z$ in the backprojection), and using a 11 GiB device RAM for the GPUs as available memory with an image of $N^3$ size with $N^2$ pixel $N$ projections, the GPUs could handle $N=17000$ before running out of memory for the projection and $N=8500$ for the backprojection. By relaxing the limits and requiring a single image slice and projection in each GPU the limit can be increased to $N=27000$. The first limit would require over 2TiB of CPU RAM to store the image, while the more permissive GPU limits would require 74 TiB of CPU RAM. We conclude that images of any of these sizes are unlikely to be used in the near future and that this code can safely claim that it can reconstruct arbitrarily large images on GPUs, even if an upper limit exists. 

The corollary is that in any realistic application the upper bound of the possible image size will be set by the CPU RAM. In addition to the limitations of storing the projections and image entirely on the CPU RAM, most iterative algorithms require auxiliary variables of the same sizes. Therefore the realistic upper bound will be set by the chosen algorithm. Using ROM as swap memory or even just writing and reading to file may increase the image size beyond the available RAM, but will result in a severe performance bottleneck due to continuous reading and writing of the auxiliary variables to ROM. We have decided not to implement this explicitly\footnote{Anyone using the code can set up their machine to use swap memory.}.   

In any case, it is possible to accelerate the general tomography code available in TIGRE if constraints are added to image/projection sizes it is expected to run. As Figure \ref{fig:pct} shows, different problem sizes would require different optimization strategies. However, we present a code that works well in a generic setting for any size and number of GPUs. Accelerating the code for a given set of GPUs with specific architecture with a specific problem size is left to the users.

On the multi-GPU side, the splitting method presented shows how single-node machines can work with multiple GPUs, however, it is limited to this computer topology. Fortunately, the strategy itself can be combined with a higher level parallelization for multiple nodes, similar to what the ASTRA toolbox provides\cite{palenstijn2017distributed}. Using the MPI protocol the code could be adapted to multiple node topologies, such as in a HPC. 
 
Results in Figure \ref{fig:pct} show further possible speed-ups. For a size of 512, the memory operations take a significant amount of the total time in the forward projection and are the main operation in the backprojection. However, most of these operations only exist due to the modular design of TIGRE. If the algorithms were to be written directly on C++/CUDA, these operations would only be needed once in the entire algorithm, instead of being used in each projection and backprojection call. Such implementation would therefore speed the algorithm even more. This means that the presented code is ready to be used for time-critical medical cases, since it can achieve reconstruction times shorter than a minute.

Computational results show that not only very large images can be successfully reconstructed regardless of the GPU, but that current image sizes can be reconstructed very fast. For a medical image size (generally smaller than $512^3$), iterative reconstruction can be achieved in less than 1 second per iteration on properly chosen hardware. As a rough comparison, the original article of TIGRE\cite{biguri2016tigre} demonstrates that a reconstruction of $512^3$ medical image using 15 iterations of CGLS can be achieved in 4min41s. With the proposed implementation the same problem can be solved in 1min01s on a single GTX 1080 Ti. With increasing image sizes and widening access to CT machines in the world, a fast yet not resource limited implementation of CT reconstruction can be of significant impact in the medical field, allowing anyone with a GPU to reconstruct using iterative algorithms.

\section{Conclusions}
 In this work we propose a projection and backprojection operation splitting strategy that allows seamlessly splitting the computation over multiple GPUs in a single node with little to no overhead. The method operates with arbitrarily large images, removing the common limit of GPU RAM size, pushing the upper bound of the maximum size that can be reconstructed in CT in single machines, and allowing significant speed-up for multiple GPU machines. Similarly, it allows reconstruction of images in machines with small GPUs. Additionally, the method allows for faster reconstruction for images that fit on GPUs. This has a clear application in medical CT, where in some cases reconstruction time is critical. With the current implementation, iterative reconstruction of medical-sized images can be performed in less than a minute. An implementation of the proposed method is integrated into the latest version of the TIGRE toolbox at \href{github.com/CERN/TIGRE}{{github.com/CERN/TIGRE}} and freely available.

\section*{Acknowledgements}
We gratefully acknowledge the support of NVIDIA Corporation with the donation of the Titan Xp GPU used for this research.  This research was supported by EPSRC grant EP/R002495/1 and the European Metrology Research Programme through grant 17IND08. The authors acknowledge the use of the IRIDIS High Performance Computing Facility, and associated support services at the University of Southampton, in the completion of this work. We thank Chris Moore Fossils and Sally Thompson for sharing the Ichthyosaur fossil data.

\FloatBarrier
\section*{References}

\bibliography{ref}

\end{document}